\documentclass[%
 preprint,
 amsmath,amssymb,
 aps,
]{revtex4-1}

\usepackage{color}
\usepackage[draft]{todonotes}   % notes showed
\usepackage{graphicx}% Include figure files
\usepackage{dcolumn}% Align table columns on decimal point
\usepackage{bm}% bold math
\begin{document}

%\preprint{APS/123-QED}

\title{Dilution of Ferromagnets via a Random Graph-based Strategy}% Force line breaks with \\

\author{Marco Alberto Javarone}
\email{marcojavarone@gmail.com}
 \affiliation{School of Computer Science, University of Hertfordshire, Hatfield AL10 9AB, UK\\
 nChain Ltd, London W1W 8AP, UK\\
Dept. of Mathematics and Computer Science, University of Cagliari, 09123 Cagliari Italy}%Lines break automatically or can be forced with \\
\author{Daniele Marinazzo}%
 \email{daniele.marinazzo@ugent.be}
\affiliation{%
Department of Data Analysis, Faculty of Psychology and Educational Sciences, University of Ghent, Ghent, Belgium
}%

\date{\today}% It is always \today, today,

\begin{abstract}
The dynamics and behavior of ferromagnets have a great relevance even beyond the domain of statistical physics.
In this work, we propose a Monte Carlo method, based on random graphs, for modeling their dilution. In particular, we focus on ferromagnets with dimension $D \ge 4$, which can be approximated by the Curie-Weiss model.
Since the latter has as graphic counterpart a complete graph, a dilution can be in this case viewed as a pruning process.
Hence, in order to exploit this mapping, the proposed strategy uses a modified version of the Erd\H{o}s-Renyi graph model. In doing so, we are able both to simulate a continuous dilution, and to realize diluted ferromagnets in one step.
The proposed strategy is studied by means of numerical simulations, aimed to analyze main properties and equilibria of the resulting diluted ferromagnets.
To conclude, we also provide a brief description of further applications of our strategy in the field of complex networks.
\end{abstract}

\maketitle

%\tableofcontents

\section{Introduction}
The study of diluted ferromagnets~\cite{galam03,ludwig01,weigt01,guerra01,contucci01,agliari01} dates back to several years ago, following two main paths sometimes overlapping, i.e. the statistical mechanics approach to lattices, and the graph theory approach to networks~\cite{barra04,montanari01}. A notable result, coming from their combination, is the modern network theory~\cite{newman01,estrada01,caldarelli01}.
In particular, the latter extends the classical graph theory to the analysis of networks characterized by non-trivial topologies and containing a big amount of nodes. So, the role of statistical mechanics is to offer methods and strategies for investigating the properties and the dynamics of these 'complex networks'~\cite{barabasi01,moreno01}.
Usually, investigations on ferromagnets are performed using the Ising model~\cite{mussardo01}, mainly because the latter constitutes a simple and powerful tool for studying phase transitions and further applications, also beyond the domain of statistical mechanics (e.g. Data Science~\cite{zecchina01} and Machine Learning~\cite{mackay01,saitta01}). Despite its simplicity, the Ising model becomes, itself, a very hard problem (not yet solved) when studied in dimensions greater than $3$.
In those cases, the Curie-Weiss~\cite{wolski01,barra05} model allows to approximate its behavior, with the advantage to be also analytically tractable (i.e. it can be exactly solved for any size of system).
As result, in some conditions, solving the Ising model might require to perform numerical simulations using Monte Carlo methods~\cite{newman02}.
For instance, the Metropolis algorithm~\cite{bhanot01} constitutes one of the early, and most adopted, strategies for simulating thermalization processes over a lattice.
This latter algorithm is based on the optimization of the Hamiltonian function representing the energy of the system.
Notably, the Hamiltonian of the Ising model reads
\begin{equation}\label{eq:ising}
H_{I}(s) = -\sum_{<i j>} J_{i,j}\sigma_i \sigma_j
\end{equation}
\noindent where the summation is extended to all the nearest neighbors $(i,j)$ in the lattice (realized with periodic boundary conditions, so actually becoming, in topological terms, a toroid).
As result, the value of the Hamiltonian~\ref{eq:ising} depends on the set $s$, i.e. the configuration of spins $\sigma$ in the lattice.
Accordingly, the two ground states of the system correspond to the spin configurations $\hat s_{+} = [+1, +1, \ldots, +1]$ and $\hat s_{-} = [-1, -1, \ldots, -1]$.
Therefore, considering a lattice with $N$ sites, and starting with a random configuration $s_x \in S$, defined as $s_x = [\sigma_1^x,...,\sigma_N^x]$, the Metropolis algorithm leads the system towards a state of equilibrium which, for a temperature $T = 0$, corresponds to one of the two ground states.
This algorithm is based on two simple steps
\begin{enumerate}
\item Randomly select a site $i$, and compute the local $\Delta E$ associated to its spin flip 
\item IF ($\Delta E \le 0$): accept the flip;\\ ELSE: accept the flip with probability $e^{\frac{-\Delta E}{kT}}$
\end{enumerate}
\noindent repeated until the equilibrium state is reached. 
We remind that $k$ and $T$, appearing in the probability shown in the step $(2)$ of the Metropolis algorithm, refer to the Boltzmann constant and to the system temperature, respectively.
In addition, the term 'local' $\Delta E$, used in the step $(1)$, indicates that the difference in energy is computed considering only the site $i$ and its nearest neighbors. Thus, in principle, some flips may increase the global energy of the whole system.
In general, the process simulated by the Metropolis algorithm takes into account the fact that the ferromagnetic interactions $J$ are \textit{quenched}, i.e. the thermalization is fast enough to allow to consider the interactions as constant.
In the opposite case, i.e. with non-constant interactions, we have different scenarios. For instance, a spin system can become glassy by introducing anti-ferromagnetic interactions (i.e. $J = -1$), or can undergo a dilution process by removing interactions (i.e. setting $J = 0$).
In this work, we focus on dilution of ferromagnets introducing a strategy, based on the Erd\H{o}s-Renyi model~\cite{erdos01}, for modeling this process.
It is worth to recall that previous investigations (e.g.~\cite{barra01,barra02,barra03,cugliandolo01}) highlighted the critical behavior of diluted ferromagnets, including for example the ergodicity breaking and the vanishing of a giant component. So, beyond providing a novel method for dilution, we give also a description of some statistical properties of the resulting system, of the dynamical processes living on it, and on potential applications.
To this end, the analyses are performed in two different conditions: for introducing the dilution strategy and studying some properties of the ferromagnets, the spin variables (i.e. $\sigma$) are considered as \textit{quenched}, while for studying thermalization processes after a dilution, the \textit{quenched} variables are the interactions $J$.
Finally, the proposed strategy and the related analyses are performed by means of numerical simulations. Beyond describing the behavior of our model, we emphasize that the achieved results allow also to envision potential applications in the area of complex networks.
The reminder of the paper is organized as follows: Section~\ref{sec:model} introduces the proposed strategy. Section~\ref{sec:results} shows results of numerical simulations. Eventually, Section~\ref{sec:conclusion} provides a description of the main findings.
\section{Modeling Dilution on Ferromagnets}\label{sec:model}
Let us consider ferromagnets of dimension $D \ge 4$, modeled via the Curie-Weiss (CW hereinafter) model. The latter is composed of $N$ sites, with a position $i$ and a spin $\sigma \pm 1$. Here, the interactions are not limited to the nearest neighbors (like in the Ising model), but are extended to all the system, i.e. every site interacts with all the others.
Accordingly, the Hamiltonian of the CW model reads
\begin{equation}\label{eq:cw}
H_{cw}(s) = \frac{-J}{2N}\sum_{i \neq j}^{N,N} \sigma_i \sigma_j
\end{equation}
\noindent with $s \in S$, i.e. combination of spins $\sigma$.
Then, like in the Ising model, the two ground states correspond to the spin configurations $\hat s_{+}$ and $\hat s_{-}$, i.e. those that minimize the value of $H_{cw}$ (Eq.~\ref{eq:cw}).
It is worth to highlight that the CW model can be represented as a complete (i.e. fully-connected) graph, where each site corresponds to a node, and each interaction to an edge.
In addition, the number of interactions is equal to $L = \frac{N \cdot (N-1)}{2}$.
The mapping from the physical object (i.e. the ferromagnet) to the mathematical entity (i.e. the graph) allows to map a dilution to a pruning process.
However, before presenting the dilution strategy developed with the framework of graph theory, we discuss the application of a more classical method, i.e. the previously mentioned Metropolis algorithm.
\subsection{Dilution by the Metropolis algorithm}
In principle, the Metropolis algorithm, and similar methods, may be used for modeling the dilution of ferromagnets.
Notably, since this algorithm modifies spins from $+1$ to $-1$, and vice versa, according to the energy difference resulting from the spin flipping, an opportune variant ---say Metropolis-like, might be used for flipping the interaction variables $J$.
In the case of spin flipping the possible values that $\sigma$ can take are $\pm 1$ whereas, in the case of interactions $J$, the latter may take three different values: $+1$ (i.e. ferromagnetic), $-1$ (i.e. anti-ferromagnetic), $0$ (i.e. removal). 
Thus, a Metropolis-like algorithm devised for flipping interactions may, in principle, generate a spin glass~\cite{parisi01,contucci02,barra09} (flipping $J$ from $+1$ to $-1$), and perform a dilution (flipping $J$ from $+1$ to $0$). In addition, both processes (i.e. from $+1$ to $-1$, and to $0$) can be combined, modeling the emergence of a diluted spin glass.
Hence, focusing on dilution, from now on, we consider only the case $J = +1 \to J = 0$.
In doing so, starting with a random distribution of spins, a Metropolis-like algorithm (M-L hereinafter) can be defined as follows:
\begin{enumerate}
\item Randomly select an interaction $J$ between two sites, and compute the local $\Delta E$ associated to its flip to $0$
\item IF ($\Delta E \le 0$): accept the flip;\\ ELSE: accept the flip with probability $e^{\frac{-\Delta E}{kT}}$
\end{enumerate}
\noindent As in the thermalization processes, the M-L strategy depends on the Hamiltonian of the system. Furthermore, one might consider also flipping of $J$ from $0$ to $+1$, i.e. modeling a kind of (edge) re-population. However, since the addition of interactions between inverse spins would increase the Hamiltonian, the actual realization of flipping $0 \to +1$ would be quite rare.
\subsection{Dilution via a Random Graphs-based Strategy}
\begin{figure*}[!h]
\centering
\includegraphics[width=16cm]{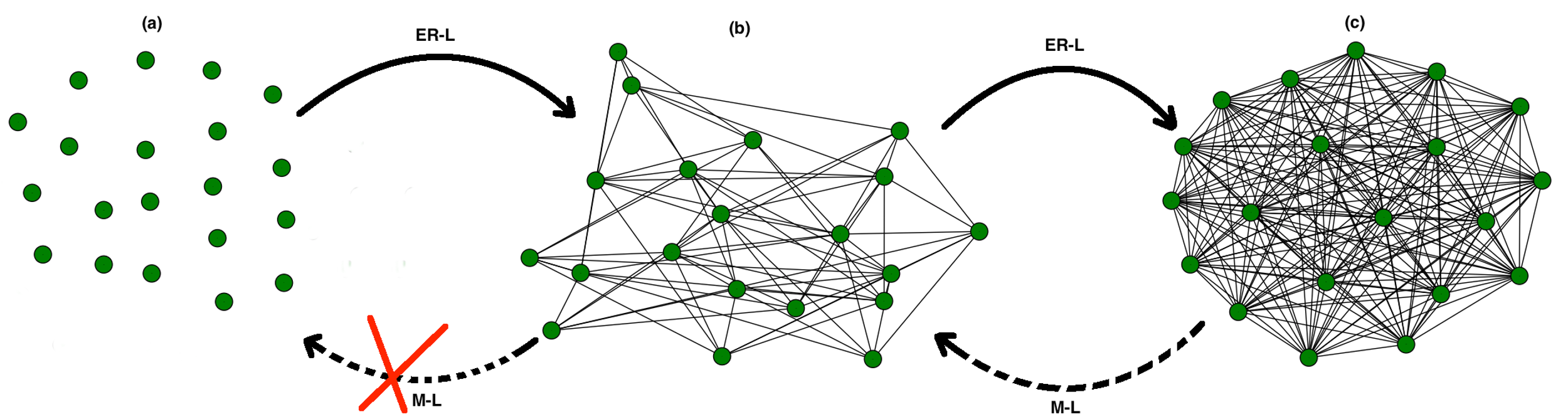}
\caption{Pictorial representation of a dilution process implemented via the ER-L strategy (continuous line) and via the M-L strategy (dotted line). Labels (i.e. \textbf{(a)},\textbf{(b)},\textbf{(c)}) indicate three different phases of the graph. Arrows indicate the direction of the process. The M-L strategy cannot lead to phase \textbf{(a)}, as indicated by the red cross on the related arrow. Here, spin values are not represented (i.e. all nodes have the same color). \label{fig:figure_1}}
\end{figure*}

As mentioned above, modern network theory and its methods are spreading in many other scientific fields. Then, it is interesting to see whether and how network theory can be useful for facing the problem of diluting ferromagnets. Notably, our work, beyond to introduce a further method for this task, allows also to prove the effectiveness of network theory in a further application. As result, the proposed model has a double valence, i.e. the process of ferromagnet dilution can be analyzed by the tools developed in network theory, and allows to envision new applications. For instance, as shown later, the subfield of community analysis can benefit from the proposed strategy.
Given this premise, we can now proceed with a brief the description of ferromagnets with the formal language of graph theory.
In general, a graph $G$ is an entity composed of two sets: $N$ (i.e. nodes) and $L$ (i.e. edges). As above reported, the maximum number of edges (i.e. $L_M$) depends on $N$. In addition, the edges can be provided with some properties, as a direction, a weight, and so on, in order to represent specific characteristics of the object their refer to (e.g. a ferromagnet, or a real network as a social network~\cite{meloni01}, a biological network~\cite{marinazzo01}, a immune network~\cite{agliari02,agliari03}, a financial network~\cite{battiston01}, and many others). 
In the proposed model, edges have no particular properties (i.e. they are indirect and unweighted), and the graph is implemented via the E-R model. The latter is realized by defining a number of nodes $N$ and a parameter $\beta$, which represents the probability of each edge to exist. Thus, the expected number of edges in an E-R graph is equal to $\mathbf{E(L)} = L_M \cdot \beta$. Notably, decreasing (increasing) $\beta$ entails to remove (add) edges in the graph.
The algorithm for generating an E-R graph is very simple:
\begin{enumerate}
\item Define the number of $N$ of nodes and the probability $\beta$
\item Draw each edge with probability $\beta$\\
\end{enumerate}
\noindent In particular, in the step $(2)$, all possible $L_M$ edges are considered.
Therefore, an E-R graph generated with $\beta = 1.0$ contains exactly $L_M$ edges (i.e. it is complete), and constitutes the graphical counterpart of the CW model previously described. 
The tuning of the parameter $\beta$ allows to represent ferromagnets with different amounts of interactions. Thus graphs generated with $\beta < 1$ (i.e. having $L < L_M$) represent diluted ferromagnets.
This last observation constitutes the base of our model, i.e. an E-R-like (ER-L hereinafter) model devised for dilution processes.
For the sake of clarity, now we provide a pictorial representation for highlighting their main differences between two mentioned strategies, i.e. M-L and ER-L ---see Figure~\ref{fig:figure_1}.
So, a quick glance to the pictorial representation allows to observe what follows: \textit{i}) the ER-L strategy starts with non-connected nodes then populates the graphs with new edges, while the M-L strategy starts with a complete graph and then removes the edges; \textit{ii}) the ER-L strategy allows to obtain more configurations than the M-L strategy, being the latter 'Hamiltonian-dependent'. In particular, once the Hamiltonian has been optimized, further actions (i.e. edge removal) have very low probability.
On the other hand, the ER-L strategy, being (partially) 'Hamiltonian-independent', allows the realization of ferromagnets with higher degree of dilution. For this reason, M-L is closer to a physical realization of a dilution than ER-L.
\subsubsection*{ER-L Strategy}
We are now ready to present the ER-L strategy in detail.
Firstly, the ER-L uses a parameter $\gamma \in [0.0, 1.0]$, representing a kind of control in the dilution process. Notably, $\gamma = 0.0$ entails the process is not controlled, while $\gamma = 1.0$ entails a fully controlled process. 
It is worth noting that, while the dilution of a ferromagnet does not require any control, being driven towards the optimization of a Hamiltonian, using a probabilistic model (i.e. the ER-L), whose dynamics depends only in part on the local energy, the so-called control parameter $\gamma$ becomes fundamental for approaching the behavior of a physical dilution.
Therefore, $\gamma$ compensates for the partial energy independence of the proposed strategy.
Accordingly, the edge probability $\beta$ is 'corrected' as follows
\begin{equation}\label{eq:beta_diff}
\beta^{\star} = \omega \cdot F_s(\omega)
\end{equation}
\noindent with $F_s$ step function and $\omega$ equal to 
\begin{equation}\label{eq:omega}
\omega = \beta + \sigma_x \sigma_y (1.0 - \beta) \gamma
\end{equation}
In doing so, $\beta^{\star} = 0$ when $\omega$ has a null or a negative value and, at the same time, the normalization condition (i.e. $\omega \le 1$) is respected for any value of $\beta$, and of $\sigma$.
Thus, varying the parameter $\beta^{\star}$, we can study the Hamiltonian of the resulting diluted ferromagnet and its behavior.
In few words, the parameter $\gamma$ makes the proposed method closer to a physical dilution, since combines $\beta$ with the contribution of the two spins involved in the interaction. For instance, from a physical point of view, an interaction between two opposite spins must be removed with a probability higher than an interaction between two equal spins. At the same time, the resulting parameter $\omega$ can take values smaller than zero, hence it cannot be directly adopted as the probability to remove an edge. As result, we introduced a 'corrected' probability $\beta^{\star}$, that takes as input any possible value of $\omega$ and has a range limited between zero and one. The degree of freedom offered by the parameter $\gamma$ allows to represent dilution processes both in physical systems, as one can do also with a more classical approach (e.g. that before described), and to consider other systems, as social networks, where further properties and mechanisms can be involved in the process. In particular, in the case of social networks, dynamical processes like dilution might consider both the node similarity (e.g. the spin) and a probabilistic process mapped to the $\beta$ parameter.
Before illustrating the results of numerical simulations, it is important to elucidate a further aspect of our investigation. As previously reported, when studying the equilibrium configuration of a spin system, the interaction variables $J$ are considered \textit{quenched}.
So, with the aim to analyze the behavior of diluted ferromagnets, the variation of $J$ must be faster than that of $\sigma$, i.e. the latter is \textit{quenched}.
Now, having defined the ER-L model, we discuss how it can be used. First, one can realize a diluted ferromagnet via ER-L, studying then its properties. Second, one can analyze the behavior of the ferromagnet during the dilution process.
While a single realization, of a diluted ferromagnet, is very similar to the realization of a graph via the E-R model, i.e. drawing edges according to the $\beta^{\star}$ probability, it is worth to explain how to implement a continuous dilution process by the ER-L model. 
To this end, let us consider a dilution as the motion of a graph $G$ over a phase space, along an axis-$\beta$. The phase space is larger for low values of $\beta$, and becomes narrower as $\beta$ increases, until it contains only one state when $\beta = 1$ (i.e. fully-connected). The lower $\beta$, the larger the number of possible realizations of $G$ with the same amount of edges (i.e. higher its entropy, see also~\cite{manlio01} for further details).
So in a continuous dilution, beyond considering the effect of $\gamma$, one can be able to move from a state, say $G(\beta_{t1})$, to a state $G(\beta_{t2})$ without losing information about the edges existing at $t1$. 
For instance, if $\beta_{t1} = 0.9$ and $\beta_{t2} = 0.8$, the ER-L must account for the removal of a density of edges equal to $0.1$, preserving the remaining structure of the graph. 
Therefore, the simple generation of a first graph with $\beta = 0.9$, and then with $\beta = 0.8$, is not allowed because the two resulting graphs are not correlated.
Thus, in the considered example, the continuous dilution process entails to move in the phase space of the graph by removing each edge with a probability much smaller than $\beta_{t2}$, in order to consider also the effect of $\beta_{t1}$.
To generalize, given $\beta_{t1}$ and $\beta_{t2}$, with $\beta_{t1} > \beta_{t2}$, if an edge $e_{ij}$ (i.e. connecting sites $i$ and $j$) belonging to the graph in the state $G_{t1}$ has to be confirmed in the state $G_{t2}$ , one cannot simply use $\beta_{t2}$ because, after the process, the edge $e_{ij}$ would be present with probability $P(e_{ij}) = \beta_{t1} \cdot \beta_{t2}$, that is obviously smaller than $\beta_{t2}$. For this reason, we need to compute the factor $\epsilon$ such that, $P(e_{ij}) = \beta_{t1} \cdot \epsilon = \beta_{t2}$. In this way, at $t2$, each edge remains in the graph with probability $\epsilon = \frac{\beta_{t2}}{\beta_{t1}}$.
In a similar fashion, we implement the inverse process, i.e. repopulating the graph with missing edges, from the state $G_{tn}$ to $G_{tn+1}$, now having $\beta_{tn} < \beta_{tn+1}$.
In particular, a new edge (always defined $e_{i,j}$) must be added to $G$ with a probability $P(e_{i,j}) = \epsilon$, coming from the following relation: $(1 - \beta_{tn})\cdot(1 - \epsilon) = (1 - \beta_{tn+1})$, so that $\epsilon = 1 - \frac{\beta_{tn}}{\beta_{tn+1}}$.
Summarizing, while a diluted ferromagnet can be realized with a single instance of the ER-L model, a continuous dilution can be implemented as follows:
\begin{enumerate}
\item Generate a graph $G (N,\beta_{0})$ and define the sampling rate for the dilution, i.e. $\Delta\beta$;
\item While $\beta_{t} > \theta$:
\item \_\_ Remove each edge in $G$ with probability $\epsilon = \frac{\beta_{t}}{\beta_{t} - \Delta\beta}$
\item \_\_ $\beta_{t} = \beta_{t} - \Delta\beta$
\end{enumerate}
The parameter $\theta$ represents the final edge probability $\beta$, i.e. the probability one should use for generating via the E-R model a graph similar to that resulting from the dilution process. $\beta_0$ corresponds to the starting value of $\beta$ for generating the initial graph, and $\beta_{t}$ corresponds to the value of $\beta$ at step $t$. 
The inverse process, i.e. the graph re-population, can be summarized as follows:
\begin{enumerate}
\item Generate a graph $G (N,\beta_{0})$ and define the sampling rate for the re-population, i.e. $\Delta\beta$;
\item While $\beta_{t} < \zeta$:
\item \_\_ Add each new potential edge in $G$ with probability $\epsilon = 1 - \frac{\beta_{t}}{\beta_{t} + \Delta\beta}$
\item \_\_ $\beta_{t} = \beta_{t} + \Delta\beta$
\end{enumerate}
As $\theta$, $\zeta$ represents the final value of $\beta$ one should use for generating a similar graph (i.e. with same statistical properties), achieved after re-population, using the E-R model. 
Moreover, we clarify that 'new potential edge' refers to the edges that can be added to the graph for making it again fully-connected, i.e. it refers only to missing edges.
Eventually, we analyze also thermalization processes (considering, after each dilution, the variables $J$ as \textit{quenched}). To this end, the system magnetization defined as
\begin{equation}\label{eq:magnetization}
M = \sum_{i} \frac{\sigma_i}{N}
\end{equation}
\noindent offers a macroscopic view on the process.
Notably, we recall that the magnetization is an order-parameter and allows both to observe the emergence of a phase transition, and to evaluate its nature (e.g. first order). In addition, it is worth emphasizing that \textit{quenched} spins, randomly initialized with a uniform distribution, entail the magnetization is on average always null (i.e. the system remains in a disordered phase).
Further analyses devised for studying the behavior of our model are introduced in the following section.
\section{Results}\label{sec:results}
The proposed model is studied by means of numerical simulations, considering ferromagnets composed of $N = 1000$ sites. In particular, we aim to obtain diluted ferromagnets with single realizations of the ER-L strategy, and to use the latter for modeling continuous dilution and re-population processes.
In addition, we analyze thermalization processes on the resulting diluted ferromagnets and, eventually, we present a potential application in the field of complex networks, i.e. in the evaluation of community stability~\cite{sole01,mougi01}.
\subsection*{Dilution via the ER-L Model}
\begin{figure*}[!ht]
\centering
\includegraphics[width=14cm]{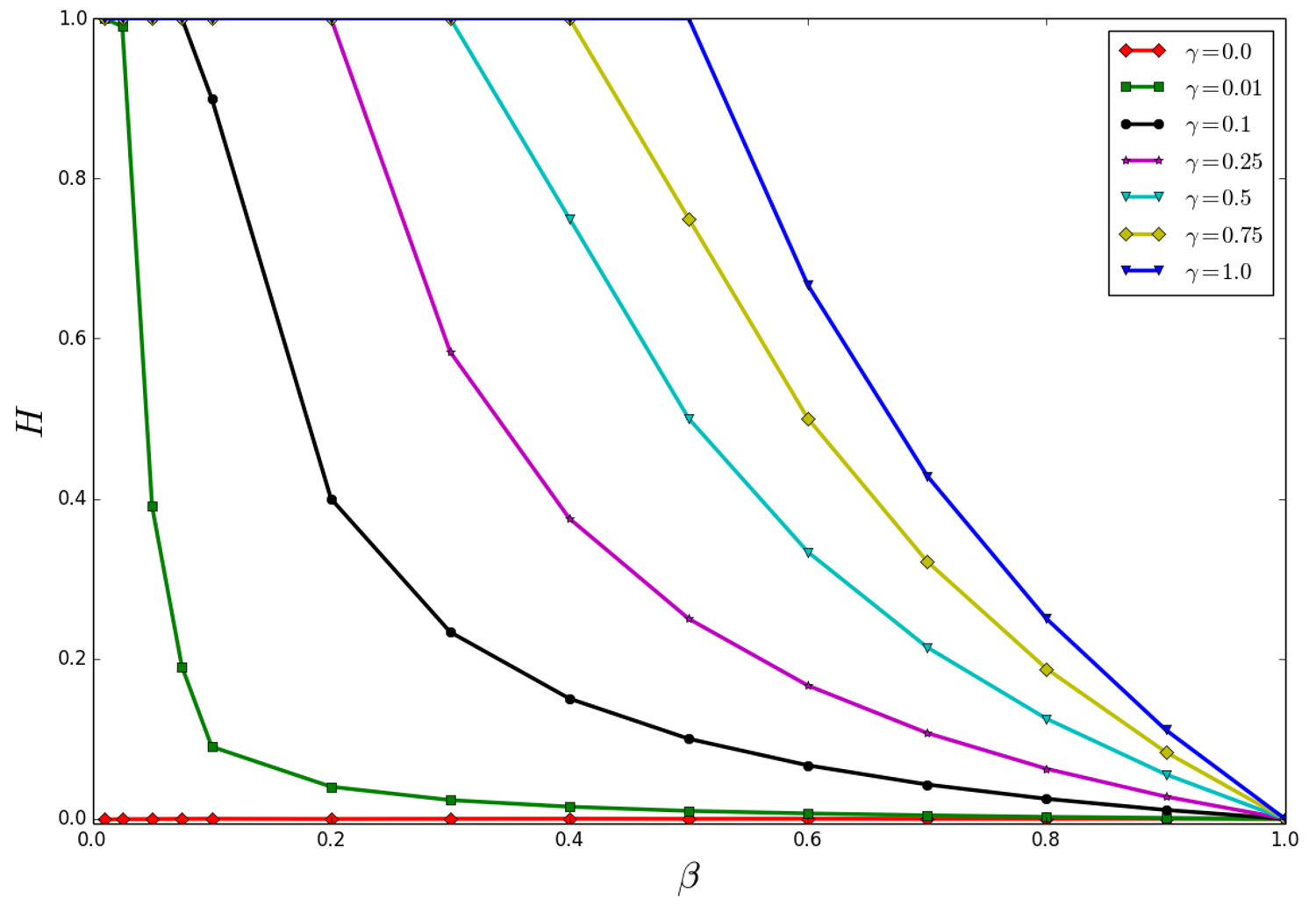}
\caption{Absolute value of the system Hamiltonian in function of the $\beta$ parameter adopted for realizing a diluted ferromagnet. The legend indicates the related value of $\gamma$ for each curve. Results have been averaged over different simulation runs. \label{fig:figure_2}}
\end{figure*}

\begin{figure*}[!h]
\centering
\includegraphics[width=14cm]{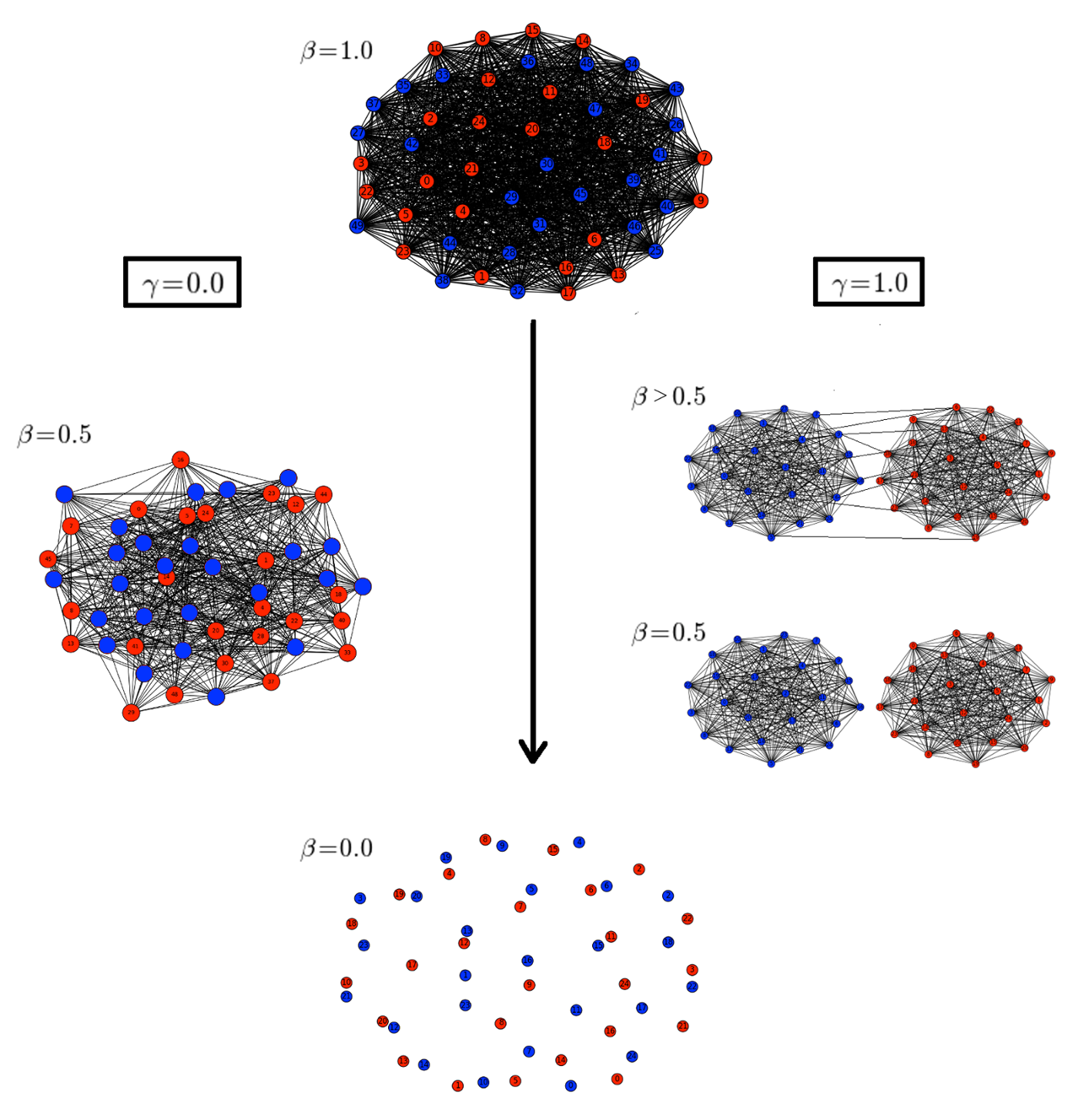}
\caption{Pictorial representation of a continuous dilution process, via ER-L, achieved in two different ways: on the left, without any control (i.e. $\gamma = 0$), and on the right with maximum control (i.e. $\gamma = 1.0$). The arrow indicates the direction of the process, starting from the top with a complete graph, up to the bottom with non-connected nodes. Along the dilution path, are shown the related graphs achieved by the different strategies previously described. In the controlled case, the graph is split in two communities for $\beta \le 0.5$, and are connected for $\beta > 0.5$. Different colors indicate different spin values (e.g. blue $\sigma = +1$, and red $\sigma = -1$). \label{fig:figure_3}}
\end{figure*}

\begin{figure*}[!h]
\centering
\includegraphics[width=14cm]{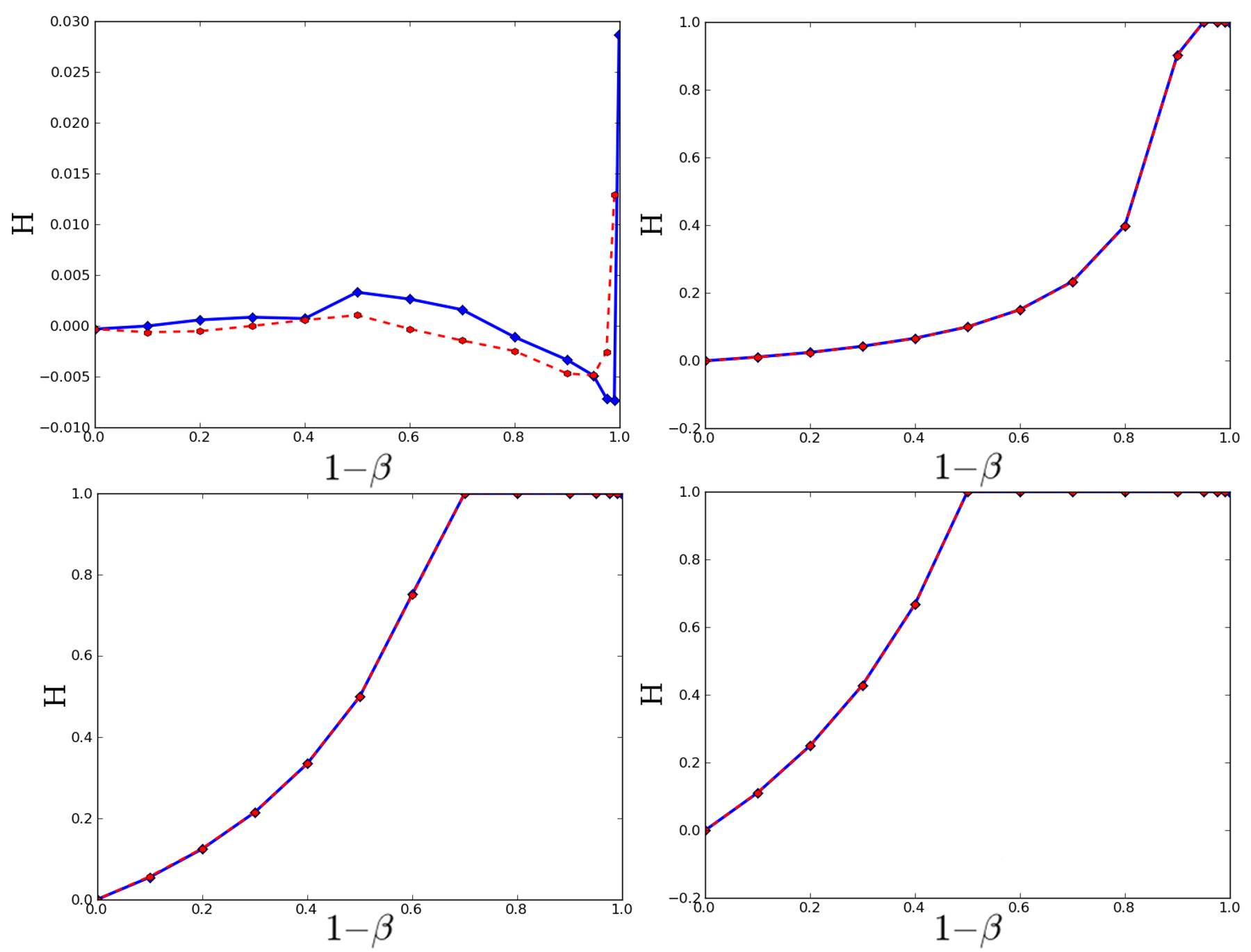}
\caption{Continuous dilution and repopulation of ferromagnets with different $\gamma$: $0.0$ (top left), $0.1$ (top right), $0.5$ (bottom left), and $1.0$ bottom right. The red curve represents the dilution, while the blue curve the repopulation process. Results have been averaged over different simulation runs. \label{fig:figure_4}}
\end{figure*}

We start considering different realizations of ferromagnets via the ER-L model, on varying $\beta$ and $\gamma$.
Figure~\ref{fig:figure_2} shows the (absolute value of) Hamiltonian $H$, normalized over the actual number of edges $L_a$, which reads
\begin{equation}\label{eq:actual_h}
H(s) = -\frac{1}{L_a}\sum_{i \neq j}^{N,N} J_{i,j}\sigma_i \sigma_j
\end{equation}
\noindent with $s$ denoting a specific spin configuration. It is important to emphasize that eq.~\ref{eq:actual_h} is normalized in order to consider only those connections that survive during the dilution process.
As expected, the Hamiltonian (eq.~\ref{eq:actual_h}) is equal to zero when there is no control in the dilution process, since interactions are removed without considering the spin of related nodes.
On the contrary, increasing $\gamma$, we observe that the Hamiltonian increases up to $1$ ---we remind that we are considering the absolute value of the Hamiltonian, so that its actual value is $-1$. 
For $\gamma > 0.0$, the maximum of $|H|$ can be reached spanning $\beta$ within well defined ranges. Notably, the latter enlarges by increasing $\gamma$. For instance, when $\gamma = 0.5$ the optimal $H$ is obtained with $0 \le \beta \le 0.3$, while when $\gamma = 1.0$ the optimal $H$ is obtained with $0 \le \beta \le 0.5$.
It is worth clarifying that we cannot study the dilution in function of $\beta^{\star}$, since its value depends on the involved spins, so consequently being potentially different from edge to edge.
This preliminary investigation constitutes an early indicator of the existence of a critical edge probability $\beta_c$, i.e. the highest value of $\beta$ to have $|H| = 1.0$.
Now, we start analyzing continuous dilution and repopulation processes. Before to show results of simulations, let us highlight that modeling a continuous dilution entails to start the process from a fully-connected graph. Therefore, considering fig.~\ref{fig:figure_3}, here the direction of a dilution corresponds to that of the M-L strategy while, obviously, a repopulation follows the inverse path.
The pictorial representation of Fig.~\ref{fig:figure_3} aims to give an overview about the continuous dilution process (via ER-L) in two conditions: non-controlled ($\gamma = 0.0$) and fully controlled (i.e. $\gamma = 1.0$). 
Provided that the starting graph and the landing one are equal in both cases, a quick glance to the pictorial allows to appreciate the influence of $\gamma > 0$. 
In particular, the intermediate graphs, between the starting and the ending ones, are related to those achieved via the ER-L method setting $\beta \approx 0.5$. Notably, in the case $\gamma = 0.0$, the graph is obtained setting $\beta = 0.5$, while in the case $\gamma = 1.0$ the plot illustrates a graph achieved with $\beta > 0.5$ and one with $\beta = 0.5$. 
Remarkably, for $\beta \le 0.5$, the resulting graph appears perfectly divided between the two communities (i.e. spins $+1$ separated from spins $-1$). Instead, for values of $\beta$ slightly higher than $0.5$, as represented in Fig.~\ref{fig:figure_3}, the two communities are connected by few edges. This observation is very important, because strongly related to thermalization processes (i.e. when the variables $J$ are taken as \textit{quenched} after the dilution step).
Numerical simulations, shown in fig.~\ref{fig:figure_4}, demonstrate that the ER-L strategy is able to dilute and to repopulate a graph, no matter the value of $\gamma$. In addition, implementing the two processes as a cycle, we did not find any form of hysteresis, i.e. dilution and repopulation cover two perfectly overlapping paths in the plot of fig.~\ref{fig:figure_4}.
Only in the case with $\gamma = 0.0$, we found an observable difference between the two paths, which can still be considered negligible.
\subsection*{Thermalization processes on diluted ferromagnets}
\begin{figure*}[!h]
\centering
\includegraphics[width=14cm]{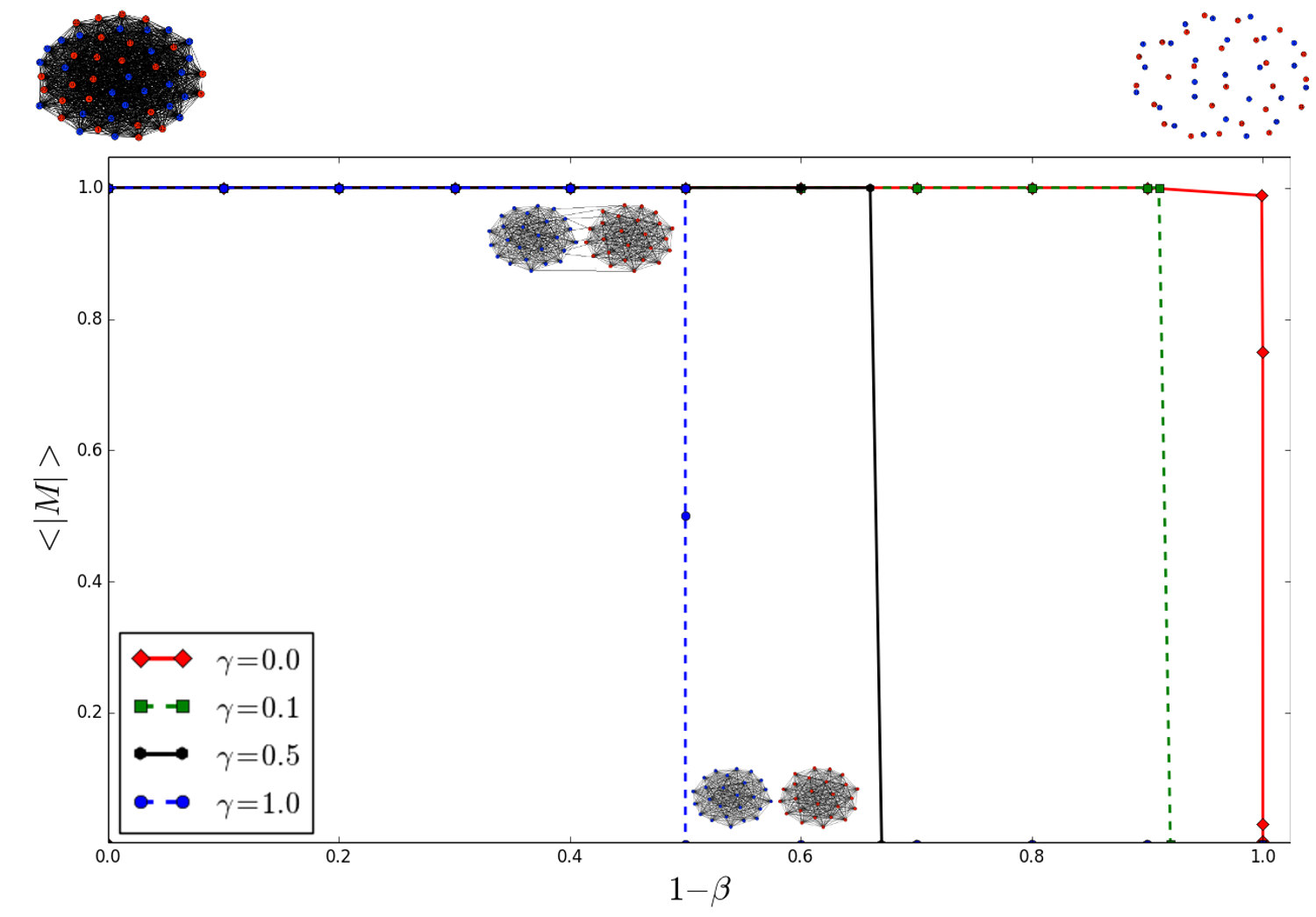}
\caption{(Absolute) Average magnetization $|M|$ in function of $1 - \beta$, for different $\gamma$, as indicated in the legend. The small networks decorate the plot, showing the dilution over the abscissa. Notably, the two inner small networks, refer to values of $\gamma = 1.0$, and to $\beta$ slightly higher and lower than $0.5$. Results have been averaged over different simulation runs. \label{fig:figure_5}}
\end{figure*}

Now, we study thermalization processes on ferromagnets diluted with different $\gamma$. To this end, ferromagnets can be diluted both implementing single realizations of the ER-L strategy (as we did here), and by performing the continuous dilution, i.e. considering the resulting graph obtained at each step. Moreover, we remind that thermalization is analyzed by studying the average magnetization (i.e. eq.~\ref{eq:magnetization}) of the system ---see Fig.~\ref{fig:figure_5}.
In all cases, it seems that the order-disorder phase transition occurring in the ferromagnet is of first order, no matter the value of $\gamma$. 
At the same time, the latter strongly affects the critical $\beta_c$.
In particular, for $\gamma = 0.0$ we found $\beta_c \sim 10^{-3}$, while for $\gamma = 1.0$ the value is smaller than $0.5 + 10^{-4}$. When $\gamma = 0$, the transition is caused by the relevant reduction of edges, so that without interactions the thermalization cannot take place. 
On the contrary, increasing $\gamma$, the order-disorder phase transition is caused initially by a combined effect of edge reduction and community separation, until $\gamma = 1.0$, where the disordered phase is reached for the emergence of two well separated, and ordered, communities having opposite spin (i.e. one with $\sigma = +1$ and one with $\sigma = -1$) ---see the inset of fig.~\ref{fig:figure_5}. In addition, we found that with $\beta = 0.5 + 10^{-5}$ and $\gamma = 1.0$, the average (absolute) value of the magnetization and the variance are equal, proving its role as the critical $\beta_c$.
Then, it is worth to further clarify an aspect shown in fig.~\ref{fig:figure_5}, i.e. the first order phase transition. Notably, when the dilution is strongly controlled (i.e. $\gamma \to 1$) the first edges to be removed are those linking nodes with opposite spin. So, that once the half of the edges is removed, only those connecting nodes with same spins survive, leading towards a total magnetization equal to zero (i.e. summing the magnetization observed in the two separated communities, which in turn reach opposite states of full order). Instead, for poorly controlled dilution, edges are removed without to consider the value of related spins, so that the transition occurs for lower values of $\beta$.
To conclude, we remind that these simulations have been performed on ferromagnets containing an equal amount of positive and negative spins.
\subsection{Community Stability}
\begin{figure*}[!h]
\centering
\includegraphics[width=14cm]{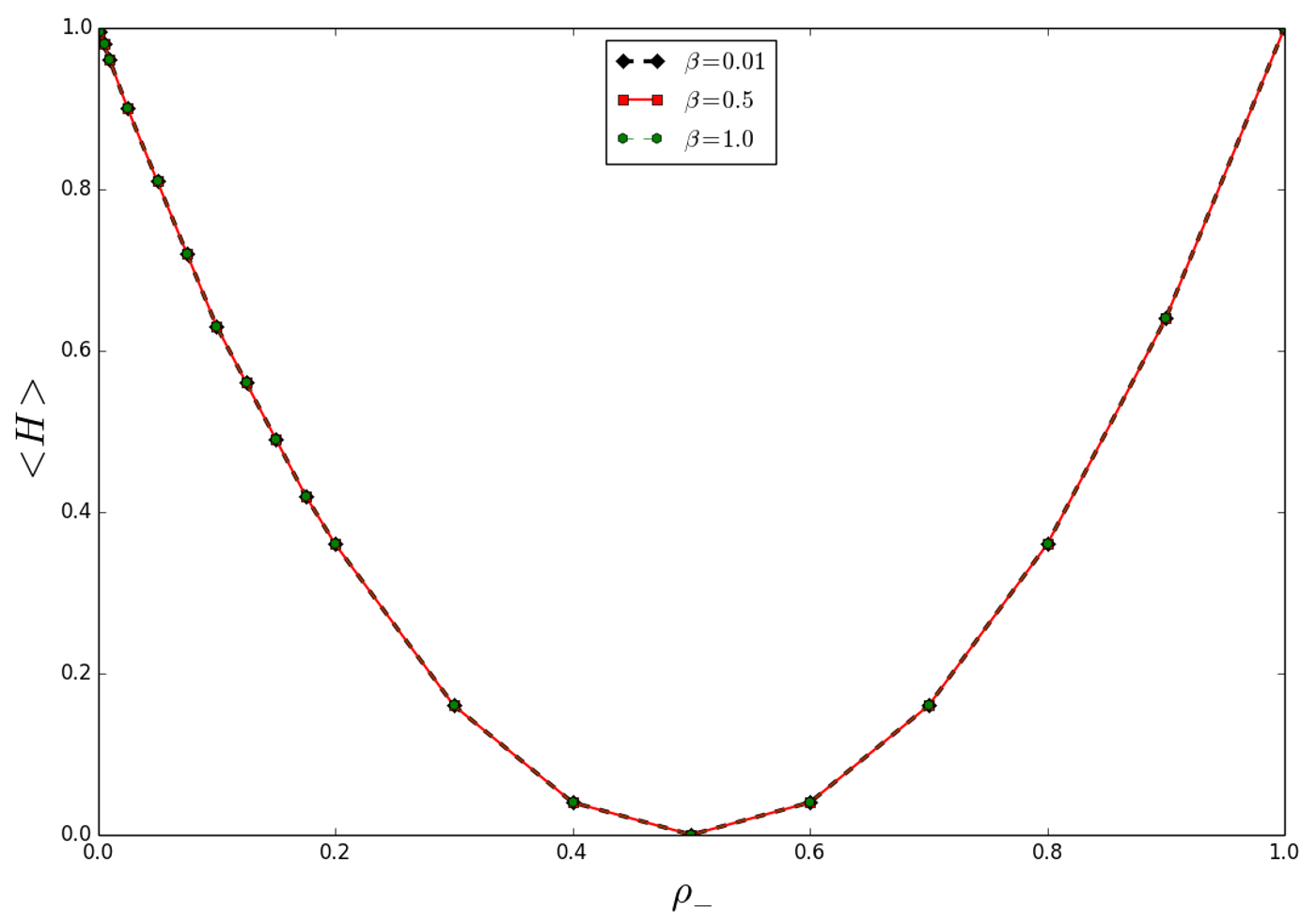}
\caption{Hamiltonian of a fully-connected community, in function of the density of negative spins. Results have been averaged over different simulation runs. \label{fig:figure_6}}
\end{figure*}

\begin{figure*}[!h]
\centering
\includegraphics[width=16cm]{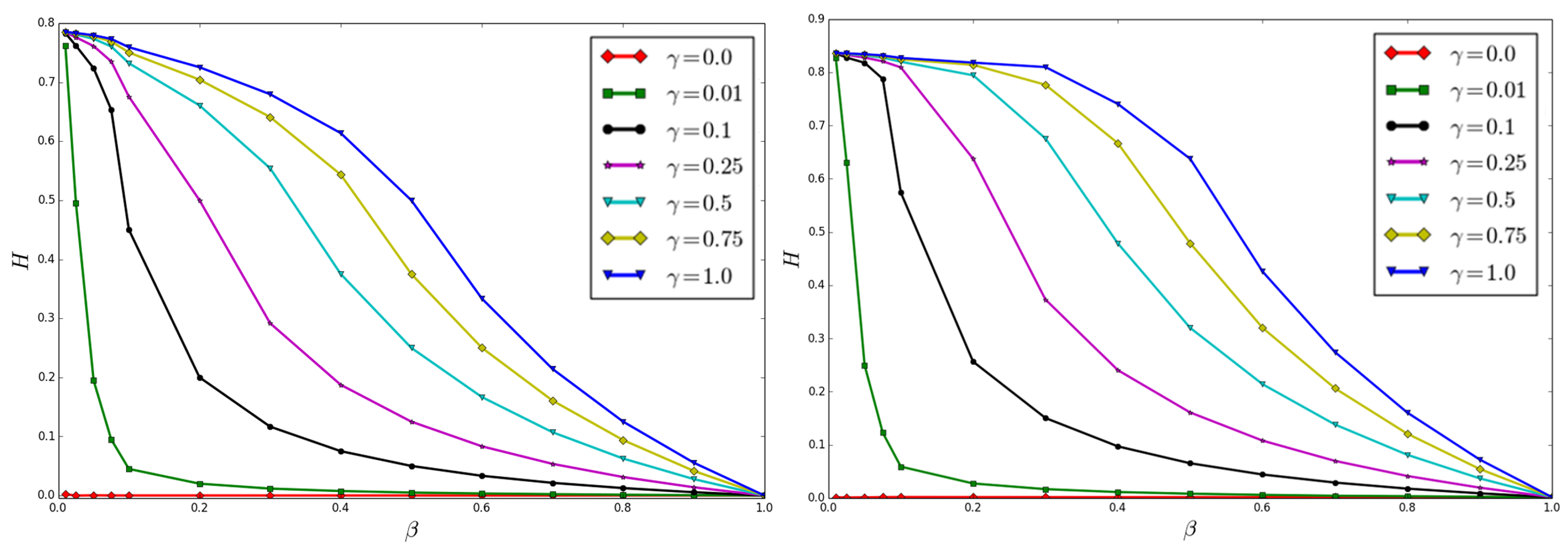}
\caption{Hamiltonian computed using the XY model. On the left, results reached considering $360$ different spin values. On the right, those reached using $4$ spin values. The legend indicates the related value of $\gamma$ for each curve. Results have been averaged over different simulation runs. \label{fig:figure_7}}
\end{figure*}

The proposed strategy aims to perform dilution processes on ferromagnets using, as reference, a well-established random graph model (i.e. the E-R model). 
The latter is widely used in the modern theory of networks for studying dynamical processes, and structural properties of complex networks.
Now, we want to evaluate if a modification of the E-R model that we introduced, i.e. the ER-L strategy, can be useful for extracting information from a complex network. In particular, we envision a potential application in the task of measuring the stability of a community, i.e. if according to the properties of its nodes, it risks to disappear after a while.
Notably, in a number of models studied in social dynamics~\cite{loreto01,galam04,bessi01,galam07}, often properties and behaviors are mapped onto binary spins~\cite{galam06}. So, in principle, one could use the Hamiltonian defined in eq.~\ref{eq:actual_h} for measuring the stability of a community~\cite{sznajd01}, i.e. the higher its $|H|$ the higher the probability that the community survives over time. In particular, the value of $|H|$ reflects the degree of similarity between the nodes connected in the same community.
Even if only the analysis of real datasets would allow to confirm the validity of this hypothesis, and then also the usefulness in the area of complex networks of the proposed strategy, our assumption is based on the simple observation that groups of individuals are more likely to cluster together when share common interests, opinions, and so on. Moreover, beyond observations of real scenarios, this mechanism is confirmed by the positive assortativity~\cite{torres01} that social networks show, i.e. individuals are more likely to interact with their likes. In addition, recalling that the value of eq.~\ref{eq:actual_h} can be related to single communities, it can be viewed as an alternative form of assortativity at community level, since the higher its value the higher the fraction of connections between similar nodes.
So, since the case with binary spins has been previously studied, even if referred to dilution processes, here we focus on two main analyses. First we study the influence of heterogeneity in a complete community, measuring how the Hamiltonian decreases while increasing the amount of nodes with different spins. Second, we study the Hamiltonian of a graph, considering the XY model~\cite{sellitto01} as reference. In doing so, we are able to represent situations where there are more than $2$ opinions (e.g.~\cite{vazquez01,galam02}), states, or behaviors. 
The first investigation considers, at the beginning, the combination named $s_{+}$, then some spins flip increasing the density of nodes with spin $-1$ (this process leads to the same results also considering the inverse case, i.e. starting with $s_{-}$ and then flipping spins to $+1$).
Results are illustrated in fig.~\ref{fig:figure_6}.
As expected, the minimum of $|H|$ is reached when the number of $+1$ spins is equal to that of $-1$ spins.
Finally, we analyzed the Hamiltonian of a community using the XY model. Figure~\ref{fig:figure_7} reports the related results, for different $\gamma$, and considering both $4$ different states, and $360$ different states. 
Here, the pairs of spins are evaluated according to the cosine similarity $\cos(\theta_a - \theta_b)$, with $\theta_a$ and $\theta_b$ representing the value of the involved spins.
We observe two main differences from the classical binary spins. In particular, using the XY model, the decrease of the Hamiltonian is smoother in the XY model than in the Ising model, where it appears less monotonous. Furthermore, the maximum value of $|H|$ is always smaller than $1.0$.
Accordingly, we note that communities are more stable (or robust to spin flipping) when there are more than $2$ states characterizing the related nodes. On the other hand, many possible states do not allow to reach a perfect stability (i.e. $|H| = 1$), exposing a community to a higher risk to disappear.
\section{Discussion and Conclusion}\label{sec:conclusion}
This work introduces a strategy, named ER-L, for modeling the dilution of ferromagnets using the framework of modern network theory. In particular, we adopt as reference the E-R graph model, since the latter, under opportune conditions, constitutes the graphical representation of the Curie-Weiss model. 
The proposed method is partially Hamiltonian-independent, i.e. while a Metropolis-like strategy can dilute a ferromagnet according to energy-based rules, the ER-L strategy depends on a probabilistic (non-physical) parameter $\beta$ and, only in part, on the local energy via a parameter $\gamma$, which represents a kind of control in the dilution.
Notably, diluting a ferromagnet can be thought as pruning a graph $G$, moving the latter in a phase space composed of all its possible realizations. The amount of edges (i.e. interactions) depends on the parameter $\beta$ of the ER-L model, so our strategy moves $G$ along the axis $\beta$. In doing so, $G$ undergoes a kind of phase transition (see also~\cite{bianconi01,bianconi02,javarone01,javarone02}), where different structures can be obtained, from sparse nodes to a complete graph.
In addition, the parameter $\gamma$ ensures that the motion along the $\beta$ axis, on the phase space, corresponds to that followed by a ferromagnet during a spontaneous dilution.
In particular, like during thermalization, a system tends to naturally reach an equilibrium state that minimizes its energy. In a similar fashion, spontaneous dilutions should lead the system towards a ground state.
Results indicate that ER-L is able to perform this task for different values of $\gamma$, depending on the considered $\beta$. In addition, we also analyzed the closed path (i.e. the cycle) from a complete graph to single nodes, and then to a complete graph by repopulating with new edges the diluted graph. 
It is worth to highlight that ER-L allows to dilute a graph also after its Hamiltonian has been minimized, while a Metropolis-like strategy, being 'Hamiltonian-dependent',  would not be able. Therefore, not all the structures obtained via ER-L have a physical meaning.
However, during a continuous dilution, we can discriminate those that appear without a physical meaning, computing the difference of the Hamiltonian between the two structures.
The related analyses have been performed considering the spin variables $\sigma$ as \textit{quenched}. So, we studied also the opposite case, i.e. after a dilution the interactions $J$ become \textit{quenched}, and the spins can flip towards an equilibrium state (see also~\cite{jmarro,torres97,marro99}). In order to study this process, and to make a relation with the parameter $\beta$, we analyze the average magnetization achieved at equilibrium, which provides an indication about the phase transitions occurring in the system ---see fig.~\ref{fig:figure_5}.
Once analyzed the outcomes of the ER-L strategy, we performed a further analysis for evaluating the opportunity to apply it to further tasks, in particular considering the measure of stability of communities in complex networks.
First we analyzed the variation of the Hamiltonian turning an ordered system to a disordered one. Then, we studied the Hamiltonian considering as reference the XY model, i.e. admitting spins with more than $2$ values.
This preliminary investigation suggests that ER-L may, in principle, be useful for evaluating the risk that a community will dissolve after a while, according to the degree of heterogeneity of its individuals (e.g. in terms of opinions, interests, and so on). In addition, we found that communities whose nodes have more possible states (e.g. opinions), never reach a perfect stability (i.e. $|H| < 1$) but can be more robust than those with binary spins to the emergence of interaction between different individuals.
Obviously, we are not taking into account all the 'social' processes that may occur in real systems, e.g. once two different individuals interact, one might imitate the other, relaxing the system.
Moreover, considering the two strategies here described, i.e. ER-L and M-L, we deem important to mention that, in principle, they might constitute also the base for developing learning algorithms~\cite{learning01}. Notably, almost all simulations have been carried on considering an equal distribution of positive and negative spins, however different combinations (i.e. patterns) might be used. Therefore, the optimization of the Hamiltonian during the dilution, in our view, even if referred to only one pattern, can be actually interpreted as a form of learning in a neural network~\cite{neural01}. On the other hand, further investigations are required for evaluating whether the proposed model may allow the graph to learn and store more than one pattern.
Finally, we remark that, in order to assess the actual usefulness of ER-L for evaluating the community stability, further investigations based on real datasets are definitely mandatory.
\\
\\
The authors declare that there is no conflict of interest regarding the publication of this paper.
\section*{Acknowledgments}
MAJ  would like to acknowledge support by the H2020-645141 WiMUST project, and to thank the mobility funds of the Faculty of Psychology and Educational Sciences of Ghent University. Authors wish to thank Adriano Barra for the priceless suggestions.


\begin{thebibliography}{99}
%
%
\bibitem{galam03} 
Galam, S., Aharony, A.:
A new multicritical point in anisotropic magnets. I. Ferromagnet in a random longitudinal field.
\emph{Journal of Physics C} \textbf{13-6} 1065 (1980)
%
\bibitem{ludwig01} 
Ludwig, A.W.W.:
Infinite hierarchies of exponents in a diluted ferromagnet and their interpretation.
\emph{Nuclear Physics B} \textbf{330-2}  639--680 (1990)
%
\bibitem{weigt01} 
Semerjian, G., Weigt, M.:
Approximation schemes for the dynamics of diluted spin models: the Ising ferromagnet on a Bethe lattice.
\emph{Journal of Physica A} \textbf{37} 5525 (2004)
%
\bibitem{guerra01} 
De Sanctis, L., Guerra, F.:
Mean field dilute ferromagnet I. High temperature and zero temperature behavior.
\emph{arxiv:0801.4940} (2008)
%
\bibitem{contucci01} 
Alberici, D., Contucci, P., Mingione, E.:
A mean-field monomer-dimer model with attractive interaction: exact solution and rigorous results.
\emph{Journal of Mathematical Physics} \textbf{55-6} 063301 (2014)
%
\bibitem{agliari01} 
Di Biasio, A., Agliari, E., Barra, A., Burioni, R.:
Mean-field cooperativity in chemical kinetics.
\emph{Theoretical Chemistry Accounts} \textbf{131-3} 1104 (2012)
%
\bibitem{barra04} 
Barra A., Agliari, E.:
Equilibrium statistical mechanics on correlated random graphs.
\emph{Journal of Statistical Mechanics: Theory and Experiment} P02027 (2011)
%

\bibitem{montanari01} 
Gerschenfeld, A., Montanari, A.:
Reconstruction for models on random graphs.
\emph{Proc. Foun. of Comp.Sci.} (2007)

\bibitem{newman01} 
Newman, M.:
The structure and function of complex networks.
\emph{SIAM Review} \textbf{45-2} 167--256 (2003)

\bibitem{estrada01} 
Estrada, E.:
The structure of complex networks: theory and applications.
\emph{Oxford University Press} (2012)

\bibitem{caldarelli01} 
Caldarelli, G.:
Scale-free networks: complex webs in nature and technology.
\emph{Oxford University Press} (2007)

\bibitem{barabasi01} 
Albert, R., Barabasi, A.L.:
Statistical mechanics of complex networks.
\emph{Review of modern physics} \textbf{74} 47--97 (2002)

\bibitem{moreno01} 
Boccaletti, S., et al.:
Complex networks: Structure and dynamics.
\emph{Physics reports} \textbf{424-4} 175--308 (2006)

\bibitem{mussardo01}
Mussardo, G.:
Statistical Field Theory: An Introduction to Exactly Solved Models in Statistical Physics.
\emph{Oxford University Press} (2010)

\bibitem{zecchina01} 
Nguyen, H.C., Zecchina, R., Berg, J.:
Inverse statistical problems: from the inverse Ising problem to data science.
\emph{arxiv:1702.01522} (2017)
%
\bibitem{mackay01} 
MacKay, D.J.C.:
Information Theory, Inference and Learning Algorithms.
\emph{Cambridge University Press} (2003)

\bibitem{saitta01} 
Saitta, L., Giordana, A., Cornuejols, A.:
Phase Transitions in Machine Learning.
\emph{Cambridge University Press} (2011)
%
\bibitem{barra05} 
Genovese, G., Barra, A.:
A certain class of Curie-Weiss models.
\emph{arxiv:0906.4673} (2009)
%
\bibitem{wolski01}
Kochmanski, M., Paszkiewicz, T., and Wolski, S.:
Curie–Weiss magnet—a simple model of phase transition Statistical Mechanics.
\emph{European Journal of Physics} \textbf{34} 1555 (2013)
%

\bibitem{newman02} 
Newman, M.E.J., Barkema, G.T.:
Monte Carlo methods in Statistical Physics.
\emph{Oxford University Press} (2001)

%
\bibitem{bhanot01} 
Bhanot, G.:
The Metropolis algorithm.
\emph{Reports on Progress in Physics} \textbf{51-3} 429 (1988)
%

\bibitem{erdos01}
P. Erd\H{o}s, P. and Renyi, A.:
On the Evolution of Random Graphs.
\emph{Publication of the Mathematical Institute of the Hungarian Academy of Sciences} 17--61 (1960)

\bibitem{barra01} 
Agliari, E., Barra, A., Camboni, F.:
Criticality in diluted ferromagnets.
\emph{Journal of Statistical Mechanics: Theory and Experiment} \textbf{10} P10003 (2008)

\bibitem{barra02} 
Agliari, E., Barra, A., Camboni, F.:
Notes on ferromagnetic diluted p-spin model.
\emph{Reports on Mathematical Physics} \textbf{68-1} 1--22 (2011)

\bibitem{barra03} 
Barra, A., Camboni, F., Contucci, P.:
Dilution robustness for mean field ferromagnets.
\emph{Journal of Statistical Mechanics: Theory and Experiment} \textbf{03} P03028 (2009)
%
\bibitem{cugliandolo01} 
Corberi, F., Cugliandolo, L.F., Insalata, F., Picco, M.:
Coarsening and percolation in a disordered ferromagnet.
\emph{arxiv:0801.4940} (2008)
%
\bibitem{parisi01} 
Mezard, M., Parisi, G., Virasoro, M.:
Spin glass theory and beyond: An Introduction to the Replica Method and Its Applications.
\emph{World Scientific} (1987)
%
%
\bibitem{contucci02} 
Contucci, P., Giardina', C.:
Perspectives on Spin Glasses.
\emph{Cambridge University Press} (2012)
%
\bibitem{barra09} 
Barra, A., Dal Ferraro, G., Tantari, D.:
Mean field spin glasses treated with PDE techniques.
\emph{EPJ-B} \textbf{86-7} 332 (2013)

\bibitem{meloni01} 
Meloni, S., Perra, N., Arenas, A., Gomex, S., Moreno, Y., Vespignani, A.:
Modeling human mobility responses to large-scale spreading of infectious diseases.
\emph{Scientific Reports} \textbf{1-62}  (2011)
%

\bibitem{marinazzo01} 
Marinazzo, D., Wu, G., Pellicoro, M., Stramaglia, S.:
Information Flow in Networks and the Law of Diminishing Marginal Returns: Evidence from Modeling and Human Electroencephalographic Recordings.
\emph{PloS ONE} \textbf{7-9} e45026 (2012)

\bibitem{agliari02} 
Agliari, E., Asti, L., Barra, A., Ferrucci, L.:
Organization and evolution of synthetic idiotypic networks.
\emph{Physical Review E} \textbf{85} 051909 (2012)

\bibitem{agliari03} 
Agliari, E., Annibale, A., Barra, A., Coolen, A.C.C., Tantari, D.:
Immune networks: multi-tasking capabilities at medium load.
\emph{Journal of Physics A: Mathematical and Theoretical} \textbf{46--33} 335101 (2013)
%
\bibitem{battiston01} 
Battiston, S., Caldarelli, G.:
Systemic risk in financial networks.
\emph{Journal of Financial Management, Markets and Institutions} \textbf{1-2} 129--154 (2013)
%
\bibitem{manlio01} 
De Domenico, M., Biamonte, J.:
Spectral entropies as information-theoretic tools for complex network comparison.
\emph{Physical Review X} \textbf{6-4} 041062 (2016)
%
\bibitem{sole01}
Montoya, J., Pimm, S., Sole', R.V.:
Ecological networks and their fragility.
\emph{Nature} \textbf{442-7100} 259--264 (2016)

\bibitem{mougi01}
Mougi, A., Kondoh, M.:
Diversity of Interaction Types and Ecological Community Stability.
\emph{Science} \textbf{337-6092} 349--351 (2012)
%
\bibitem{loreto01}
Castellano, C. and Fortunato, S. and Loreto, V.:
Statistical physics of social dynamics.
\emph{Rev. Mod. Phys.} \textbf{81-2} 591--646 (2009)

\bibitem{galam04}
Galam, S.:
Sociophysics: a review of Galam models.
\emph{International Journal of Modern Physics C} \textbf{19-03} 409--440 (2008)

\bibitem{bessi01} 
Del Vicario, M., et al.:
The spreading of misinformation online.
\emph{PNAS} \textbf{113-3} 554--559 (2016)

\bibitem{galam07} 
Galam, S., Javarone, M.A.:
Modeling Radicalization Phenomena in Heterogeneous Populations.
\emph{PloS ONE} \textbf{11-5} e0155407 (2016)

\bibitem{galam06} 
Galam, S.:
Contrarian deterministic effects on opinion dynamics:'the hung elections scenario'.
\emph{Physica A} \textbf{333} 453--460 (2004)

\bibitem{sznajd01}
Sznajd-Weron, K. and Sznajd, J.:
Opinion Evolution in Closed Community.
\emph{International Journal of Modern Physics C} \textbf{11-6} 1157 (2000)

\bibitem{torres01}
Johnson, S., Torres, J., Marro, J., Mu\~noz, M.A.:
Entropic Origin of Disassortativity in Complex Networks.
\emph{Physical Review Letters} \textbf{104-10} 108702 (2010)
%
\bibitem{sellitto01} 
Berthier, L., Holdsworth, P.C.W., Sellitto, M.:
Nonequilibrium critical dynamics of the two-dimensional XY model.
\emph{J.Phys.A.: Math Gen} \textbf{34} 1805 (2001)

\bibitem{vazquez01} 
Vazquez, F., Krapivsky, P.L., Redner, S.:
Constrained opinion dynamics: freezing and slow evolution.
\emph{J.Phys.A:Math Gen.} \textbf{36} 61--68 (2003)

\bibitem{galam02} 
Gekle, S., Peliti, L., Galam, S.:
Opinion dynamics in a three-choice system.
\emph{EPJ-B} \textbf{45-4} 569--575 (2005)

\bibitem{bianconi01} 
Bianconi, G., Barabasi, A.L.:
Bose-Einstein condansation in complex networks.
\emph{Physical Review Letters} \textbf{86-24} 5632 (2001)
%
\bibitem{bianconi02} 
Ferretti, L., Mamino, M, Bianconi, G.:
Condensation and topological phase transitions in a dynamical network model with rewiring of the links.
\emph{Physical Review E} \textbf{89-4} 042810 (2014)
%
\bibitem{javarone02} 
Javarone, M.A., Armano, G.:
Quantum Classical Transition in Complex Networks.
\emph{Journal of Statistical Mechanics: Theory and Experiment} P04019 (2013)
%
\bibitem{javarone01} 
Javarone, M.A.:
Fermionic networks: Modeling adaptive complex networks with fermionic gases.
\emph{Int. J. Mod. Phys. C} \textbf{27-02} 1650021 (2016)

\bibitem{jmarro} 
Marro, J., Dickman, R.:
Nonequilibrium phase transitions in lattice models.
\emph{Cambridge University Press} (2005)

\bibitem{torres97} 
Torres, J.J., Garrido, P.L., Marro, J.:
Neural Networks with Fast Time-Variation of Synapses.
\emph{Journal of Physica A: Mathematical and General} \textbf{30-22} 7801 (1997)


\bibitem{marro99} 
Marro, J., Torres, J.J., Garrido, P.L.:
Neural Networks in which Synaptic Patterns Fluctuate with Time.
\emph{Journal of Statistical Physics} \textbf{94-5} 837--858 (1999)

%
\bibitem{learning01} 
Zdeborova, L., Krzakala, F.:
Statistical physics of inference: Thresholds and algorithms.
\emph{Advances in Physics} \textbf{65-5} (2016)
%
\bibitem{neural01} 
Agliari, E., et al.:
Multitasking associative networks.
\emph{Physical Review Letters} \textbf{109} 268101 (2012)
%
\end{thebibliography}
\end{document}